\documentclass[aps,showpacs,prl,twocolumn,superscriptaddress]{revtex4} 

\usepackage{amssymb}
\usepackage{graphpap}
\usepackage[dvips]{graphicx}
\usepackage[dvips]{graphics}
\usepackage{color}

\begin{document}

\title{Electron-Hole Crossover in Graphene Quantum Dots}

 \author{J. G\"uttinger}
 \affiliation{Solid State Physics Laboratory, ETH Zurich, 8093 Zurich, Switzerland}
  \author{C. Stampfer}
 \affiliation{Solid State Physics Laboratory, ETH Zurich, 8093 Zurich, Switzerland}
  \author{F. Libisch}
  \affiliation{Institute for Theoretical Physics, Vienna University of Technology, 1040 Vienna, Austria, EU}
  \author{T. Frey}
   \affiliation{Solid State Physics Laboratory, ETH Zurich, 8093 Zurich, Switzerland}
     \author{J. Burgd\"orfer}
  \affiliation{Institute for Theoretical Physics, Vienna University of Technology, 1040 Vienna, Austria, EU}
  \author{T. Ihn}
 \affiliation{Solid State Physics Laboratory, ETH Zurich, 8093 Zurich, Switzerland}
   \author{K. Ensslin}
 \affiliation{Solid State Physics Laboratory, ETH Zurich, 8093 Zurich, Switzerland}

\date{ \today}
 
\begin{abstract}
We investigate the addition spectrum of a graphene quantum dot in the vicinity of the electron-hole crossover as a function of perpendicular magnetic field.  Coulomb blockade resonances of the 50~nm wide dot are visible at all gate voltages across the transport gap ranging from hole to electron transport. The magnetic field dependence of more than
50 states displays the unique complex evolution of the diamagnetic spectrum of a graphene dot from the low-field regime to the Landau regime with the $n=0$ Landau level situated in the center of the transport gap marking the electron-hole crossover. The average peak spacing in the energy region around the crossover decreases with increasing magnetic field. In the vicinity of the charge neutrality point we observe a well resolved and rich excited state spectrum.

\end{abstract}

\pacs{73.22.-f, 72.80.Rj, 73.21.La, 75.70.Ak}  
\maketitle

It is an important goal in quantum dot physics to understand the quantum mechanical energy spectra and the corresponding orbital and spin states of these man-made artificial atoms. This knowledge is necessary for designing and operating quantum dots in regimes relevant for particular applications, for example, for the implementation of qubits. It has been shown in various material systems~\cite{Drexler94,Tarucha96,Ciorga00,JaHe04} that usually the measured spectra can only be easily understood, and related to theoretical models, in the few-electron (or hole) limit. In recent research on quantum dots made from graphene~\cite{Geim07,sta08a,pon08,gue08,sch09}, ground and excited states~\cite{sch09} have been observed in the Coulomb blockade regime, but attempts to relate the observations to theoretical spectra remain dissatisfactory so far, mainly
because the number and character (electron or hole) of charge carriers in the dots are unknown and
fingerprints of the graphene-specific two-dimensional linear dispersion have not been found.

Here we characterize the electron-hole crossover in high-quality graphene quantum dots (QD). We present experimental and theoretical results for the evolution of a large number of resonances 
in a graphene quantum dot near the charge neutrality 
point in a magnetic field from the low-field regime to the 
regime of Landau level formation. The addition spectrum 
displays several intricate features specific to graphene, 
among them formation of the lowest ($n = 0$) Landau level 
at high B-fields. In the following we exploit the magnetic field dependence of the manifold of low energy states to approximately pin down the electron-hole crossover point.

\begin{figure}\centering
\includegraphics[draft=false,keepaspectratio=true,clip,%
                   width=0.8\linewidth]%
                   {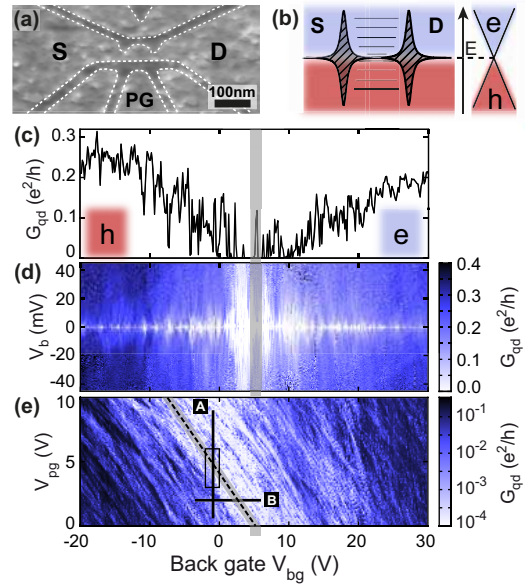}                   
\caption[FIG1]{(Color online) (a) SFM image of the graphene QD device with source (S) and drain (D) leads, and lateral plunger gate (PG). 
(b) Schematic level structure of the QD device highlighting the constriction induced tunneling barriers and the electron-hole crossover.
(c) Back gate (BG) dependent conductance at $V_{b} = 4$~mV, where transport is tuned from the hole (h) to the electron (e) regime.
(d) Conductance as function of bias and BG voltages exhibiting the energy gap of the tunneling barriers [see panel (b)]. 
(e) Source-drain current as function of PG and BG at $V_{b} = 100~\mu$V. The slope of the dashed line corresponds to the relative PG/BG lever arm ([A] and [B] mark regions studied in more detail below).
}
\label{transport}
\end{figure}

\begin{figure*}\centering
\includegraphics[draft=false,keepaspectratio=true,clip,%
                   width=0.95\linewidth]%
                   {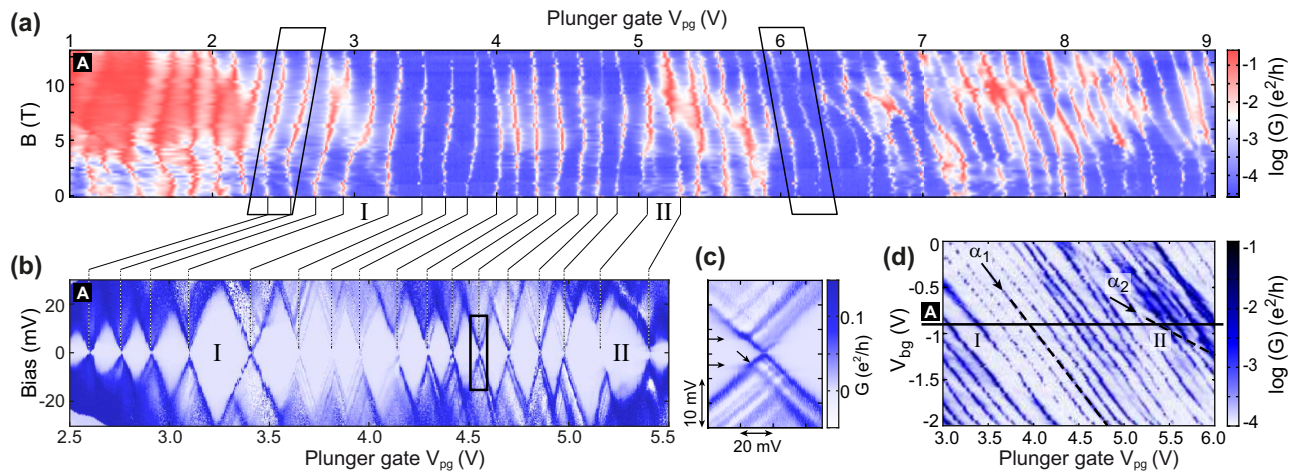}                   
\caption[FIG2]{(Color online). (a) Evolution of Coulomb blockade resonances in perpendicular magnetic field. The conductance is plotted as function of plunger gate voltage $V_{pg}$ and magnetic field $B$ ($V_{b}$ = 100~$\mu$V). 
(b) Coulomb diamond measurements at $B =~$0~T. 
(c) Close-up of the box in (b) 
highlighting the rich excited state spectra. (d) Close-up of Fig.~1(e) (see box therein) showing (Coulomb) resonances (with lever arms $\alpha_{1,2}$; dashed lines) as function of PG and BG ($V_{b} = 100~\mu$V). The solid line [A] at $V_{bg}$ = -0.9 V marks the regime where (a) and (b) have been measured [see also [A] in Fig.~1(e)]. 
} 
\label{experiment}
\end{figure*}

The $\approx 50$~nm wide and $\approx 80$~nm long quantum dot [scanning force microscope (SFM) image, Fig.~1(a)] is connected to source (S) and drain (D) via two 25~nm wide and 10~nm long
constrictions, acting as tunneling barriers [Fig.~1(b)].
The dot and the leads can be tuned by an in-plane graphene plunger gate (PG) and the highly doped silicon substrate is used as a back gate (BG).
The sample fabrication is summarized in Ref.~\cite{sta08a}.
Here, the minimum feature size of the graphene
nanodevice has been reduced compared to Ref.~\cite{sta08a} by a (reactive ion) etching step based on a patterned 45~nm thick resist layer.

Measurements were performed in a dilution refrigerator at a base temperature of 90~mK and an electron temperature of $\approx$~200~mK  (determined from the width of conductance resonances in the Coulomb blockade regime). The source-drain bias $V_b$ is symmetrically applied across the dot and the conductance is measured using low-frequency lock-in techniques.

A large-scale back-gate characteristic of the device conductance (taken at $V_{b} = 4$~mV) features a crossover from the hole (h) to the electron (e) regime [see Fig.~1(c)]. Within the transition region (0~V~$< V_{bg} <$ 10~V) conductance is strongly suppressed. This so-called transport gap~\cite{sta09,tod09,liu09} is caused by local tunneling barriers at the two (25~nm) constrictions [see Fig.~1(b)] responsible for the QD device functionality~\cite{sta08a}. We find a corresponding
bias voltage gap exceeding 40~meV around the charge neutrality point [see Fig. 1(d)], in good agreement with earlier work~\cite{han07}.
The QD charge can be selectively tuned by the nearby PG, allowing tunneling spectroscopy of dot states by reducing the influence of parasitic resonances in the constrictions~\cite{sta09,mol09}. 
Fig.~1(e) shows 
low-bias conductance measurements as function of BG and PG.
The transport gap (bright area) measured
as function of the back gate shifts with a slope given by the relative PG/BG lever arm $\alpha_{pg,bg}^{(gap)} \approx 1.3$. 

Much more fine structure in the $V_{pg}$ - $V_{bg}$ parameter plane appears on finer voltage scales [Fig.~2(d), which is a closeup of Fig.~1(e), see box therein].
We observe a sequence of essentially straight tilted lines
signifying Coulomb blockade resonances of the QD.
The corresponding relative PG/BG lever arm is given by $\alpha_1 = \alpha_{pg,bg}^{(qd)} = 1.27$ [left dashed line in Fig.~2(d)], comparable to the influence of the PG on the gap ($\alpha_{pg,bg}^{(gap)}$).

In Fig.~2(b) we show corresponding Coulomb-blockade diamond
measurements, i.e., measurements of the differential
conductance ($G=dI/dV_b$) as function of bias voltage
$V_b$ and $V_{pg}$ for fixed $V_{bg}=~-$0.9~V [solid line [A] in Figs.~2(d) and 1(e)].
We observe strong fluctuations of the addition energy ranging from $13-32$~meV. 
A close-up of the boxed region in Fig.~2(b) [see Fig.~2(c)] shows a well resolved and rich excited state spectrum with corresponding co-tunneling
onsets (see arrows). 
The energies of the lowest excited states at positive and negative bias are 3.1~meV and 2.3~meV (arrows). These energies are slightly higher than what has been reported in~\cite{sch09}, consistent with the smaller size of this QD. 

\begin{figure*}[t]\centering
\includegraphics[draft=false,keepaspectratio=true,clip,%
                   width=1.0\linewidth]%
                   {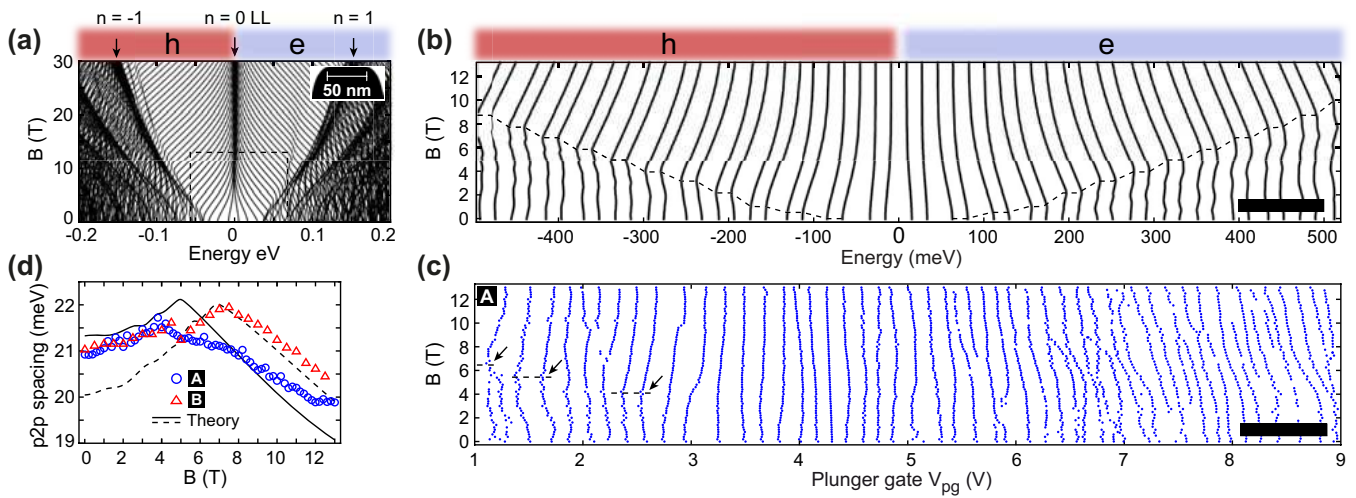}                   
\caption[FIG3]{(Color online). (a) 
Calculated levels of a graphene QD (see inset) as function of magnetic field. Arrows mark the lowest Landau level ($E_0$), i.e., the electron-hole crossover at $n = 0$. 
(b) Close-up of dashed box in panel (a), where spin, Zeeman splitting and a constant charging energy of~18~meV are included.
 (c) Experimental data. The peak positions are extracted from the measurements shown in Fig.~2(a). 
The scale bar corresponds to 100~meV. 
 (d) Average peak-to-peak spacing as function of $B$ for the two different regimes [A] (circles) and [B] (triangles) highlighted in Fig.~1(e), solid [A] and dashed [B] are calculated from (b).} 
\label{theory}
\end{figure*}

In Fig.~2(a) we show about 50 Coulomb resonances [including those discussed in Figs.~2(b)-(d)] as a function
of a magnetic field B perpendicular to the sample plane. 
The measurement has been taken in two steps with an overlapping part (below $V_{pg} = 7~$V), reproducible except 
for a small, compensated shift of $\Delta V_{pg} = 50~$mV. 
The full $V_{pg}$ range is highlighted by the vertical line [A] in Fig.~1(e). We tune from hole to electron transport since we cover the whole transport gap [see Figs.~1(c)-(e)]. 

The evolution of Coulomb resonances in the (perpendicular) magnetic field shows a common trend at lower PG voltages (see e.g., the box at $V_{pg} \approx 2.6~$V) to bend for increasing B-field towards higher energies (higher $V_{pg}$). In contrast, we find for higher PG voltage regimes the opposite trend (see e.g., the box at $V_{pg} \approx 6~$V), where resonances tend to shift to lower energies for increasing B-field.
This overall pattern is disturbed by additional
features such as localized states, regions of multi-dot behavior and strong amplitude modulations
probably due to constriction resonances. For example, we observe around $V_{pg} = 3~$V (marked with I) 
a weakly coupled state crossing the resonances to the left at B~=~5 and 9~T. At B~=~0~T this state is not visible in transport but leads to a large diamond and an extended peak-to-peak spacing in Fig.~2(d) [I in Figs.~2(b),(d)]. We interpret the weak magnetic field dependence of this state and its low visibility in transport as a manifestation of a localized state as probably caused by disorder.
Additionally, we observe several level crossings and splittings in the region of $V_{pg} = 5-6~$V (starting with II). 
This is consistent with the presence of additional resonances following a different PG/BG lever arm [$\alpha_2 \approx 0.53$ see upper right of Fig.~2(d)]. Mixing of these states with dot states also leads to deformed 
Coulomb-blockade diamonds [see II in Fig.~2(b)].

From Fig.~2(a) the peak positions are extracted and plotted in [Fig.~3(c)]. 
Reproducible peak shifts for all Coulomb blockade peaks 
were compensated based on a linear fit to the resonance at 4.25~V. 
In addition the weakly coupled state around $V_{pg} = 3$~V is removed.

To elucidate the evolution of the Coulomb resonances with
magnetic field, we performed numerical calculations of
the eigenenergies of a graphene QD in a perpendicular B-field. 
We use a third-nearest neighbor tight-binding approximation to
simulate a 50$\times$80~nm graphene QD [see inset in Fig.~3(a)]
containing $\approx$~150~000 carbon atoms (for details see Ref.~\onlinecite{lib09}). The magnetic field is included by a Peierls phase factor.
To eliminate contributions from edge states localized at the zigzag-boundaries~\cite{Fujita} we employ an on-site potential of $\pm 2 eV$ on all carbon atoms with
dangling bonds \cite{BeenBoundary}. 

The evolution of the large-scale spectrum of the QD as function of
$B$ [Fig.~3(a)] exhibits several remarkable features unique to
graphene. Near zero magnetic field, the density of states (DOS) is low
near $E=0$ and increases with increasing $|E|$, $\rho \propto |E|$,
reflecting the DOS of the Dirac cone. With increasing B, the spectrum
undergoes a morphological change such that the DOS eventually reflects the
asymptotic Landau levels (LL) of the two-dimensional Dirac equation,
$E_n=\mathrm{sgn}(n)\sqrt{2e\hbar v_{\mathbf F}^2|n|B}$. Most prominently, the
$n=0$ level with $E_0=0$ becomes field-independent and can serve to
uniquely mark the electron-hole crossover point. Accordingly, single
particle levels in the vicinity of the Dirac point converge from both
negative and positive energies towards $E=0$ nearly
symmetrically. This pattern resembles the analytical results for
circular dots~\cite{sch08, rec09}. For larger $|E|$ the evolution of
the spectrum as function of B involves a multitude of avoided
crossings giving rise to a condensation near ``ridges'' which follow
the B dependence of the higher Landau levels $E_n \propto \sqrt
{B|n|}$. These large scale features are found to be robust against
changes in the edge disorder and a moderate concentration of
additional charged impurities~\cite{lib09b}. 
The mean single-particle level spacing $\delta E \approx 4$ meV is significantly smaller than the charging energy $E_C \approx 18$ meV estimated from the Coulomb diamond measurements in Fig.~2(b). To emulate the spectrum of the experimental QD, we therefore include $E_C$ as well as spin (i.e.~Zeeman) splitting [see Fig.~3(b)]. We observe an additional remarkable feature: 
The magnetic field induces level crossings of the single particle levels, being visible as small fluctuations of the peak positions [barely visible in Fig.~3(b)]. In analogy to Ref.~\cite{Ciorga00} no crossings appear beyond the filling factor $\nu = 2$ boundary (ridge of the first Landau level). In contrast to GaAs quantum dots this leads to pronounced ``kinks'' in the peak positions [dashed line in Fig.~3(b)] because in graphene the n=0 Landau level does not shift in energy with increasing magnetic field. This behavior is a unique consequence of the interplay between the linear two dimensional (2D)-dispersion of graphene, and the finite-size dot. 


Comparing the numerical data and the measurement [Figs.~3(b) and (c)] we find the same qualitative trend of states running towards the center ($E_0$). Some kinks in the experimental data may be attributed to level crossings with the filling factor $\nu = 2$ boundary 
[see arrows in Fig.~3(c)]. Beyond the data presented here, these kinks in the
magnetic field dependence of Coulomb resonances may be used in future
experiments to identify the few-electron regime in graphene. 

The convergence of levels towards $E_0$ can be quantified by the
nonmonotonic evolution of the mean level spacing [Fig.~3(d)]. Measured
peak-to-peak spacing within $V_{pg} = 2.4-4.9$~V [at $B = 0~$T in
Fig.~3(c)], are compared with calculated level spacings for $-315$~meV $ < E < -15$~meV [at $B = 0~$T in Fig.~3(b)]. The energy range for the comparison is obtained by converting the PG and BG voltage into energy using the lever arms $\alpha_{pg} = 0.125$, $\alpha_{bg} = 0.1$, and allowing for an energy offset of 530~meV 
obtained by comparison of Figs.~3(b),(c). In the experiment we find a maximum at $B~ \approx 4$~T followed by a decrease for increasing $B$, consistent with the theoretical expectation (solid line), reflecting the
convergence of single-particle levels towards $E_0$. The slower decrease in the experiment may be attributed to the presence of additional localized states as described above. 
These findings are confirmed by additional measurements taken in a different regime
sweeping the BG across the charge neutrality point [see horizontal line [B] in Fig.~1(e)].  
We find the same qualitative behavior as in Fig.~2(a) (not shown). In Fig.~3(d) (triangles) we plot the mean peak-to-peak spacing of resonances $-1.3 < V_{bg} < 1.7$~V. 
The data points have been shifted down by -2.5~meV for better comparison with theory. By using the same energy offset (530~meV) and lever arms as before the corresponding energy interval $-415$~meV $ < E < -115~$meV is obtained. Also for this regime the shape of the measured mean energy spacing agrees well with that obtained by theory. The maximum is now shifted to higher magnetic field ($B~\approx 7$~T) as expected for an energy window shifted to larger $|E|$.

By comparison with theory, we assign the location of the transition region between hole and electron transport to the states ranging from $V_{pg} = 4.3~$V to 5.5~V ($V_{bg} = -0.9$~V). Using the relative lever arm $\alpha_{pg,bg}^{(qd)} \approx 1.27$ [see dashed line in Fig.~1(e)] this region is converted to an effective back gate voltage scale and plotted as a gray shaded box in Figs.~1(c)-(e). The extracted crossover region is in good agreement with the center of the transport gap as expected if the QD and the constriction regions have comparable doping. 

In conclusion, we have performed detailed studies of the electron-hole crossover in the addition spectrum of a graphene quantum dot in a perpendicular magnetic field. 
We observe in the evolution of the spectrum as function of the magnetic field unique features of both the linear dispersion and of the edge confinement of the graphene QD. One prominent example is the development of the graphene specific $E = 0$ Landau level around the charge neutrality point. The overall tendency of states converging towards the zero-energy Landau level is accompanied by kinks which we attribute to crossings with higher Landau levels. 
Coulomb diamond measurements in this transition regime show a rich spectrum of excited states and co-tunneling. 
These measurements open the way for more detailed studies of the electron-hole transition including a better understanding of the addition spectra and spin states in graphene quantum dots.

The authors wish to thank J.~Seif, P.~Studerus, C.~Barengo, T.~Helbling and S.~Schnez for help and discussions. 
Support by the ETH FIRST Lab, the Swiss National Science Foundation and NCCR nanoscience are gratefully acknowledged. F. L. and J. B. acknowledge support by the FWF (Austria).

 \newpage

\end{document}